\documentclass[final,times,fleqn,
5p,twocolumn,
sort&compress,merge
]{elsarticle}

\usepackage{amsmath, amssymb}
\usepackage{slashed}
\usepackage{nicefrac}
\usepackage{graphicx}
\usepackage[subrefformat=parens,labelformat=parens]{subfig}
\usepackage{mathrsfs}
\usepackage{microtype}

\usepackage[colorlinks, citecolor=blue, linkcolor=blue, urlcolor=blue]{hyperref}

\renewcommand{\vec}[1]{\ensuremath{\mathbf{#1}}}
\newcommand{\im}{\mathrm{i}}
\newcommand{\diff}{\mathrm{d}}
\newcommand{\smat}{\mathcal{S}}
\renewcommand{\smat}{\mathscr{S}}
\newcommand{\rate}{\mathcal{R}}
\renewcommand{\rate}{R}
\newcommand{\Efield}{\mathcal{E}}
\newcommand{\compl}{\bar}

\newcommand{\weak}{\text{w}}
\newcommand{\strong}{\text{s}}
\renewcommand{\weak}{1}
\renewcommand{\strong}{2}

\newcommand{\abs}[1]{\left|\,{#1}\,\right|}
\newcommand{\vecprod}[2]{{#1}\!\cdot\!{#2}}
\newcommand{\scalar}[3][\mu]{#2_#1 #3^#1}

\newcommand{\sci}[2]{{#1}{\times}{10^{#2}}}
\newcommand{\unit}[1]{~\text{#1}}

\newcommand{\figref}[1]{Fig.\,\ref{fig:#1}}
\newcommand{\subfigref}[1]{Fig.\,\protect\subref*{fig:#1}}

\newcommand{\equref}[1]{Eq.\,\eqref{eq:#1}}

\newcommand{\figlab}[1]{\label{fig:#1}}

\newcommand{\equlab}[1]{\label{eq:#1}}
\newcommand{\seclab}[1]{\label{sec:#1}}

\begin{document}
\begin{frontmatter}

\title{Nonperturbative Bethe--Heitler pair creation\\ in combined high- and low-frequency laser fields}

\author[duess,mpik]{Sven Augustin\corref{corresponding}}
\cortext[corresponding]{Corresponding author}
\ead{sven.augustin@mpi-hd.mpg.de}

\author[duess,mpik]{Carsten M\"uller}
\ead{carsten.mueller@tp1.uni-duesseldorf.de}

\address[duess]{Institut f\"ur Theoretische Physik I, Heinrich-Heine-Universit\"at D\"usseldorf, Universit\"atsstr. 1, 40225 D\"usseldorf, Germany}
\address[mpik]{Max-Planck-Institut f\"ur Kernphysik, Saupfercheckweg 1, 69117 Heidelberg, Germany}

\journal{Physics Letters B}
\date{\today}

\begin{abstract}
The nonperturbative regime of electron-positron pair creation by a relativistic proton beam colliding with a highly intense bichromatic laser field is studied. The laser wave is composed of a strong low-frequency and a weak high-frequency mode, with mutually orthogonal polarization vectors. We show that the presence of the high-frequency field component can strongly enhance the pair-creation rate. Besides, a characteristic influence of the high-frequency mode on the angular and energy distributions of the created particles is demonstrated, both in the nuclear rest frame and the laboratory frame.
\end{abstract}


\end{frontmatter}

\section{Introduction}

In the presence of very strong electromagnetic fields, the quantum vacuum can decay into electron-positron pairs \cite{Sauter,*heisenberg,Greiner}. Pair creation in a constant electric field was studied in detail by Schwinger \cite{Schwinger}, employing the then newly established methods of quantum electrodynamics (QED). The resulting pair-creation rate has the form $\rate \sim \exp(-\pi \, \nicefrac{\Efield_\text{cr}}{\Efield})$, where $\Efield$ is the applied field and $\Efield_\text{cr} = \nicefrac{m^2}{e}$ the critical field of QED. Here, $e = \abs{e}$ and $m$ are the positron charge and mass, respectively, and relativistic units with $\hbar = c = 1$ are used. The Schwinger rate has a manifestly nonperturbative character due to its non-analytic field dependence. Therefore, it is of fundamental importance for our understanding of nonperturbative quantum field theories. An experimental verification has so far been prevented, though, by the huge value of $\Efield_\text{cr} = \sci{1.3}{16}~\nicefrac{\text{V}}{\text{cm}}$, which is not accessible in the laboratory yet.

The recent progress in high-intensity laser devices has strongly revived the interest in Schwinger pair creation.%
\footnote{%
For recent work on the Schwinger effect, see \cite{ringwald,*ruffini,*bulanov-schwinger,*dumlu-dunne-schwinger,*hebenstreit,*kohlfuerst, ehlotzky-review,*dipiazza-review} and references therein.} %
In particular, various schemes have been proposed to facilitate its observation. Among them is a dynamically assisted variant where a rapidly oscillating electric field is superimposed onto a slowly varying electric field \cite{schuetzhold,*dunne}. The total rate for pair creation in this field combination was shown to be strongly enhanced, while preserving its nonperturbative character. This interesting prediction has led to a number of subsequent investigations. Currently, the momentum distributions of pairs created by the dynamically assisted mechanism are under active scrutiny \cite{orthaber,fey,li-zi-liang-boson}.

Pair-creation phenomena resembling the Schwinger effect can also occur when a beam of charged particles collides with a counterpropagating intense laser beam. For instance, pair creation has been theoretically predicted in relativistic proton-laser collisions via a strong-field version of the Bethe--Heitler effect (see, e.g., \cite{cm-circ, Avetissian, Sieczka, Milstein, Kuchiev, tomueller2, fillion}). When the laser frequency $\omega$ is relatively small, so that the dimensionless parameter $\xi \equiv \nicefrac{e \Efield}{m \omega}$ becomes large, $\xi \gg 1$, the total rate for this process adopts the form $\rate \sim \exp(-2 \sqrt{3} \, \nicefrac{\Efield_\text{cr}}{\Efield'})$ \cite{Milstein}. Here, $\Efield' = (1+\beta)\gamma \, \Efield$ denotes the laser field amplitude transformed to the rest frame of the proton, which is assumed to be sub-critical: $\Efield' \ll \Efield_\text{cr}$. The proton Lorentz factor $\gamma$ and reduced velocity $\beta$ have been introduced here. The similarity with the Schwinger rate is due to the fact that the characteristic length scale of pair 
creation in this parameter regime is much shorter than the laser wavelength, so that the field appears as quasi-static during the process.

An experimental setup to observe the strong-field Bethe--Heitler process could, in principle, be realized by utilizing the highly relativistic proton beam of the Large Hadron Collider (LHC) at CERN ($\gamma \sim 7000$) in conjunction with a counterpropagating high-intensity laser beam ($\Efield \sim 10^{11}~\nicefrac{\text{V}}{\text{cm}}$). In fact, a very similar scheme was employed to observe pair creation by multiphoton absorption at the Stanford Linear Accelerator Center, where a $46 \unit{GeV}$ electron beam colliding with an intense optical laser pulse triggered the pair creation \cite{e144-1997}.
This experiment, however, did not probe the Schwinger regime of pair creation because the field intensity was somewhat too low ($\xi \approx 0.3$). Instead, the measured pair-creation rate showed a power-law dependence on the field strength.%
\footnote{%
However, an onset of nonperturbative signatures has been observed in \cite{e144-1997};
see also \cite{reiss-trident,*hu-trident}.
} %
Note that, a rate of the form $\rate \sim \xi^{2n}$, with $n$ denoting the number of absorbed laser photons, follows from a consideration within the $n$-th order of perturbation theory, which is valid for $\xi \ll 1$.

Also for the strong-field Bethe--Heitler process, a dynamically assisted variant has been studied recently \cite{dipiazza-tunnel,dipiazza}.
There, it has been shown that the total pair-creation rates in relativistic proton-laser collisions may be strongly enhanced when a weak high-frequency component is superimposed onto an intense optical laser beam. Moreover, similar enhancement effects have been obtained for pair creation by the strong-field Breit--Wheeler process \cite{Jansen} and in spatially localized fields \cite{jiang-grobe,*jiang-grobe2}.

In this letter, we study strong-field Bethe--Heitler pair creation in a laser field consisting of a strong low-frequency and a weak high-frequency component. An $\smat$-matrix calculation within the framework of laser-dressed QED employing Dirac--Volkov states is performed. Besides total pair-creation rates, our method allows us to calculate angular and energy distributions of the created particles that are not accessible to the polarization operator approach used in \cite{dipiazza-tunnel,dipiazza}.
The nonperturbative interaction regime with $\xi \approx 1$ is analyzed both in the laboratory and the nuclear rest frame.
We note that Bethe--Heitler pair creation in bichromatic laser fields has been studied recently, with a focus on interference and relative phase effects arising in the case of commensurable frequencies \cite{krajewska-phase,Roshchupkin,augustin,*augustin2}.

It is interesting to note that analogies of the Schwinger effect exist in other areas of physics. They have been demonstrated in various systems such as graphene layers in external electric fields \cite{allor-graphene,*klimchitskaya-graphene}, ultracold atom dynamics \cite{szpak-cold2},
and light propagation in optical lattices \cite{dreisow-waveguide}. In these systems, the transition probability between quasiparticle and hole states shows Schwinger-like properties.

\section{Theoretical Framework} \seclab{calc}

Employing the $\smat$-matrix formalism of electron-positron pair creation in combined laser and Coulomb fields, we write the transition amplitude in the nuclear rest frame as \cite{cm-circ, Avetissian, Sieczka}
\begin{equation} \equlab{smat}
\smat = 
\im \, e \!
\int \compl\Psi_{p_-,s_-}^{(-)} \, \gamma_0 A_\text{N}(r) \, \Psi_{p_+,s_+}^{(+)} \, \diff^4 \! x.
\end{equation}
Electron and positron are created with free momenta $p_\pm$ and spins $s_\pm$, where the subscripted sign denotes their respective charge. They are described by relativistic Volkov states $\Psi_{p_\pm,s_\pm}^{(\pm)}$ taking their interaction with a plane-wave laser field fully into account.
In our case, this laser field is a superposition of two independent modes with the combined vector potential:
\begin{equation}
\vec{A}_\text{L} = \sum_{i=1,2} \vec{a}_i \cos(\eta_i),
\end{equation}
wherein the phase variables $\eta_i = \scalar{x}{k_i} = \omega_i t - \vecprod{\vec{k}_i}{\vec{r}}$ are defined via the wave vectors $k_i = \omega_i \kappa$,
where $\kappa = \left( 1, 0, 0, 1 \right)$. The individual amplitude vectors $\vec{a}_i$ are chosen to be perpendicular and their absolute value is given by the dimensionless intensity parameters
\begin{equation} \equlab{xi}
\xi_i = 
\frac{e}{m} 
\frac{\abs{\vec{a}_i\!}}{\sqrt{2}}.
\end{equation}
The nuclear Coulomb field $A_\text{N}(r) = \nicefrac{Ze}{r}$, with the nuclear charge number $Z$, is treated in the lowest order of perturbation theory.

The $\smat$-matrix may be evaluated by expanding the occurring periodic functions into Fourier series. The thereby introduced series summation indices $n_i$, with $i=1$ or $2$, may be understood as numbers of photons taken from the respective laser mode $i$.
This expansion allows to perform the four-dimensional integration from \equref{smat} analytically by using the Fourier transform of the Coulomb potential and a representation of the $\delta$-function for the integral in space and time, respectively \cite{bjorken-drell}:
\begin{equation} \equlab{int4d}
\int \! \diff^4 \! x ~ A_\text{N}(r) \,
\exp\!{
\left(
\im \,
\scalar{x}{Q}\right)}
=
\frac{4 \pi Ze}{Q^2}
~
2 \pi
\,
\delta(Q_0),
\end{equation}
where
$Q = q_+ + q_- - n_1 k_1 - n_2 k_2$
is the momentum transfer to the nucleus and
\begin{equation}
q_\pm
=
p_\pm
+
\frac{e^2 \overline{\vec{A}_\text{L}^2}}{2 \scalar{{\kappa}}{p_\pm}} \, \kappa
\end{equation}
are the laser-dressed momenta.
Note that, by definition of $Q_0$, the arising $\delta$-function ensures energy conservation.

The fully differential pair-creation rate is obtained by summing the square of the amplitude $\smat$ over the final spin states:
\begin{equation} \equlab{rate}
\diff^6 R = 
\frac{1}{T}
\sum_{s_+, s_-}
\abs{\smat}^2
\frac{\diff^3 q_-}{(2\pi)^3}
\frac{\diff^3 q_+}{(2\pi)^3},
\end{equation}
where the time interval $T$ has been introduced.
For incommensurable frequencies, the squared $\smat$-matrix adopts the form
\begin{equation}
\abs{\smat}^2 \sim \sum_{n_1, n_2} \abs{M_{n_1, n_2}}^2 \delta(Q_0) \, T,
\end{equation}
with the matrix element $M_{n_1, n_2}$ containing the Fourier series coefficients.

For the following discussion, it will be particularly convenient to rewrite \equref{rate} by introducing \emph{partial rates} distinguished by a single summation index, such that
\begin{align} \equlab{rate-sum-n1}
\diff^6 R &= \sum_{n_1} \diff^6 R_{n_1}, \\
\diff^6 R_{n_1}
&\sim
\sum_{n_2}
\sum_{s_+, s_-}
\abs{M_{n_1, n_2}}^2
\delta(Q_0)
\frac{\diff^3 q_-}{(2\pi)^3}
\frac{\diff^3 q_+}{(2\pi)^3}.
\end{align}
The $\delta$-function introduced in \equref{int4d}, allows to perform one integration analytically. In order to gain total rates or rates differential in a single coordinate the necessary remaining integrations are then calculated numerically.

\section{Results} \seclab{results}

\subsection{Total Pair-Creation Rates}

In the following we shall apply our formalism to pair creation by a relativistic nuclear beam and a bichromatic laser field, which is composed of a high-frequency low-intensity (i.e., weak) mode $(\omega_\weak \sim 2 m$, $\xi_\weak \ll 1)$ and a low-frequency high-intensity (i.e., strong) mode $(\omega_\strong \ll 2 m,$ $\xi_\strong \sim 1)$. Such a field may be achieved experimentally by the following combination in the laboratory frame (primed):
The assisting photon from the weak field with $\omega'_\weak = 70 \unit{eV}$ may be provided by an extreme-ultraviolet (XUV) source, e.g., one based on High-Harmonic Generation (HHG) \cite{zhang-hhg} or a Free-Electron Laser (FEL) \cite{ackermann-fel}.
For the strong field, the $1064 \unit{nm}$  infrared (IR) light emitted by an Nd:YAG (neodymium-doped yttrium aluminum garnet) laser may be frequency doubled in order to deliver $\omega'_\strong = 2.33 \unit{eV}$. This corresponds to a frequency ratio of $\nicefrac{\omega_\weak}{\omega_\strong} \approx 30$.
In combination with a proton beam of $\gamma \lesssim 7460$, as currently provided by the LHC \cite{LHC-TDR}, a setup of technology available today is feasible.
Specifically, we will first examine the case of $\gamma = 6600$, with the resulting photon energies in the nuclear rest frame $\omega_\weak = 924 \unit{keV}$ and $\omega_\strong = 30.76 \unit{keV}$, as well as $\xi_\weak = \sci{1.45}{-8}$.
Note that, the nuclear rest frame energy of the assisting mode is indicated by the notion of a \emph{$\gamma$-assisted} process.
Furthermore, the parameters of the assisting mode correspond to an intensity of approximately $10^7~\nicefrac{\text{W}}{\text{cm}^2}$ in the laboratory frame. We point out that this is a moderate value given the fact that present-day HHG and FEL sources may reach field intensities of the order of $10^{13}~\nicefrac{\text{W}}{\text{cm}^2}$ \cite{mashiko-hhg-intense} and $10^{18}~\nicefrac{\text{W}}{\text{cm}^2}$ \cite{young-fel-intense}, respectively.

\begin{figure}[t]
\centering
\subfloat[][$\gamma = 6600$, $\omega_\weak = 924 \unit{keV}$, and $\omega_\strong = 30.76 \unit{keV}$]{\figlab{assist924}%
\includegraphics[width=\columnwidth]{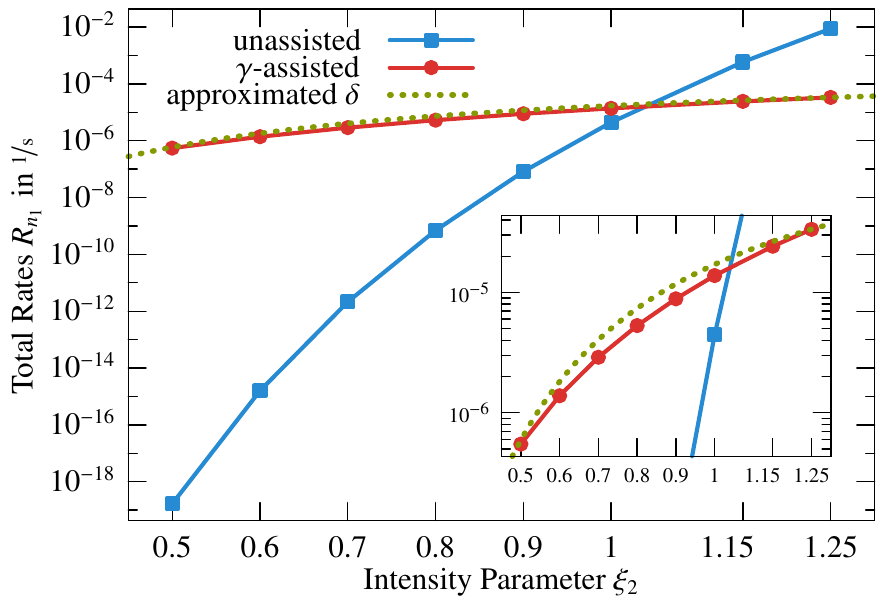}}
\\
\subfloat[][$\gamma = 7000$, $\omega_\weak = 980 \unit{keV}$, and $\omega_\strong = 32.62 \unit{keV}$]{\figlab{assist980}%
\includegraphics[width=\columnwidth]{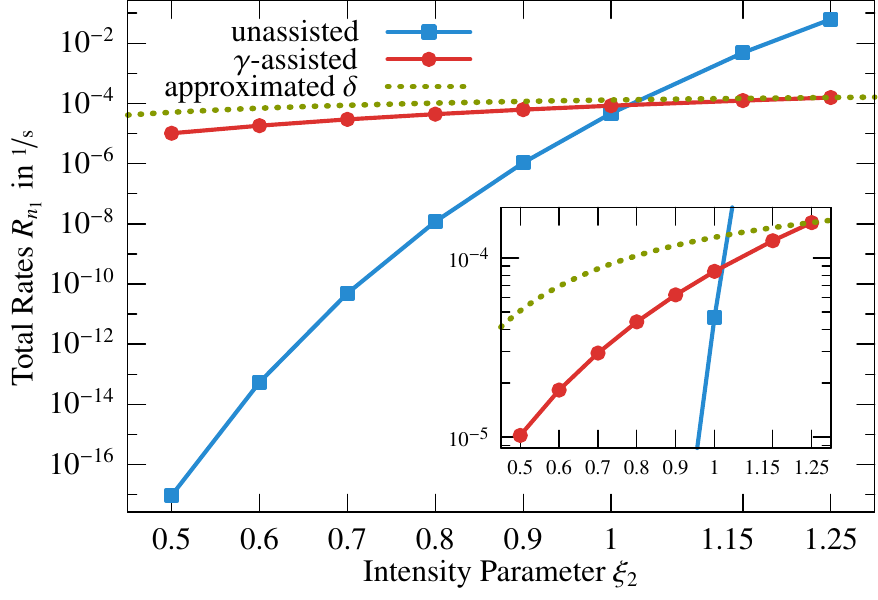}}
\caption{\figlab{assist}%
Total pair-creation rates of the unassisted and the $\gamma$-assisted process in the nuclear rest frame
--
For the latter, the result obtained here is compared to the analytical counterpart extracted from \,\cite{dipiazza-tunnel}, as can be seen in the insets.
For intensities below $\xi_\strong \approx 1$, the assisted process dominates. Above this value, the situation inverses.
In both figures, $\xi_\weak = \sci{1.45}{-8}$ is used.
}
\end{figure}

We shall begin with the comparison of strong-field Bethe--Heitler pair creation with and without the assistance of a single high-energetic photon. The respective total pair-creation rates $R_{n_1}$ are shown in \subfigref{assist924} for the nuclear rest frame. The corresponding rate in the laboratory frame is given by $R'_{n_1} = \nicefrac{R_{n_1}}{\gamma}$.
Even though the depicted intensity range of the strong mode is $\xi_\strong \approx 1$, the total rates of both processes show a Schwinger-like dependence on this parameter:
\begin{equation} \equlab{schwinger}
\rate \sim \exp\left(-\frac{C}{\xi_\strong}\right).
\end{equation}
However, the coefficient $C$ differs for the two cases. In the assisted process $C$ is smaller, leading to an almost flat curve, while in the unassisted case a steep increase is visible.

Taking the absolute height into account, the influence of the assisting photon on the total pair-creation rate becomes apparent: As long as the intensity of the highly intense laser is below $\xi_\strong \approx 1$, the assisted process shows a strong enhancement over the unassisted counterpart. However, from $\xi_\strong \approx 1$ onwards the situation inverses and the unassisted process becomes dominant.%
\footnote{%
Pair creation by the weak mode ($\xi_\weak = \sci{1.45}{-8}$, $\omega_\weak = 924 \unit{keV}$) alone, without contribution from the strong mode (hence, for $\xi_\strong = 0$), is dominated by the absorption of two $\omega_\weak$-photons.
We find a pair-creation rate of $\sci{4.6}{-17} \nicefrac{1}{\text{s}}$, which is much smaller than that of the respective dominating process in \subfigref{assist924}. The same is true for $\omega_\weak = 980 \unit{keV}$, with a slightly larger rate of $\sci{6.1}{-17} \nicefrac{1}{\text{s}}$.
\label{fn:weakonly}
} %
Note that, the value of $\xi_\strong$ where this crossing occurs may be influenced by adjusting $\xi_\weak$, on which the unassisted process does not depend and the assisted process depends quadratically. Therefore, a relative scaling between the two graphs is possible.

In a similar scheme, total rates of $\gamma$-assisted pair creation have been investigated fully analytically in \cite{dipiazza-tunnel}. Particularly, Eq. (9) therein (here quoted in a form adapted to our notation),
\begin{equation}\equlab{formula9}
\rate \sim \exp\left(-\frac{2\sqrt{2}}{3\zeta}\right),
\end{equation}
with the parameters
\begin{align}
\zeta &= \frac{\chi}{\delta^{\nicefrac{3}{2}}}, \equlab{param-zeta}
\quad \text{where} \quad
\chi = \frac{1}{\sqrt{2}} \frac{\omega_\strong}{\omega_\weak} \xi_\strong, \\
\delta &= \left( \frac{2m}{\omega_\weak} \right)^{\mkern-4mu 2} - 1 \approx \frac{2m - \omega_\weak}{m}, \equlab{param-delta}
\end{align}
allows for a comparison of the derived dependence on the strong field intensity $\xi_\strong$ with our numerically integrated result. This comparison is done by scaling the exponential via a proportionality factor such that the values at $\xi_\strong = 1.25$ coincide.
However, we note that \equref{formula9} was obtained under the assumption that $\xi_\strong \gg 1$. Even though this requirement is not met for the results shown in \subfigref{assist924}, where $\xi_\strong \sim 1$, the additional graph -- for the approximated $\delta$ -- shows very good agreement.

Besides this requirement of large intensities, there are two additional constraints demanded for the applicability of \equref{formula9}: The two parameters given in \equref{param-zeta} and \equref{param-delta} should be $\ll 1$.
For the parameter set above, their values are both relatively small ($\approx 0.2$) indeed. However, if we keep the laser combination in the laboratory fixed and increase $\gamma$ to $7000$, leading to photon energies in the nuclear rest frame of $\omega_\weak = 980 \unit{keV}$ and $\omega_\strong = 32.62 \unit{keV}$, we end up with $\zeta \approx 0.9$.
This expectedly leads to the significant deviation between our result and that obtained using \equref{formula9} clearly visible in \subfigref{assist980}.

Had we chosen a $\gamma$ lower than the initial $6600$, the results would also have deviated from the prediction of \equref{formula9}, as then the parameter $\delta$, indicating the remaining energy gap that has to be overcome by absorption of photons from the strong laser mode $(i = 2)$, would have grown to and eventually overcome unity.
Based on our results, we can establish an empirical extension to \equref{formula9} valid for this regime. It has been obtained by expressing \equref{formula9} in the form
\begin{equation}\equlab{reformula9}
\rate \sim \exp\left(-\frac{4}{3} \frac{1}{\omega_\strong} ~ \omega_\weak \left(2 - \frac{\omega_\weak}{m}\right)^{\nicefrac{3}{2}}  \frac{1}{\xi_\strong^d} \right),
\end{equation}
where we have introduced an empirical fitting parameter $d$. Note that, for $d$ set to unity, \equref{reformula9} is identical to \equref{formula9} if the approximated expression for $\delta$ from \equref{param-delta} is employed.
Then we separately identified dependences on the two photon energies via fits to our results from a wide range of parameters, $800 \unit{keV} \le \omega_\weak \le 900 \unit{keV}$ and $10 \unit{keV} \le \omega_\strong \le 100 \unit{keV}$ in the nuclear rest frame, with according functions extracted from \equref{reformula9}. From this we could gain that for $d \approx 0.8$ the exponential scaling works well when $\delta$ is not very small. While this is only a small deviation from the Schwinger-like form of \equref{schwinger}, it can strongly impact the quantitative value of the predicted rates.

Note that, for the regime where the parameter $\zeta$ is not very small, i.e., for energies of the assisting photon just below the pair-creation threshold like in \subfigref{assist980}, no constant empirical value for $d$ could be determined, as the dependence on $\xi_\strong$ changes rapidly with the precise value of the two parameters. This regime shall be further investigated in the subsequent section.

\subsection{Angular- and Energy-Differential Spectra}

\begin{figure}[t]
\centering
\subfloat[][Nuclear Rest Frame]{\figlab{spectra-nuc-theta}%
\includegraphics[width=\columnwidth]{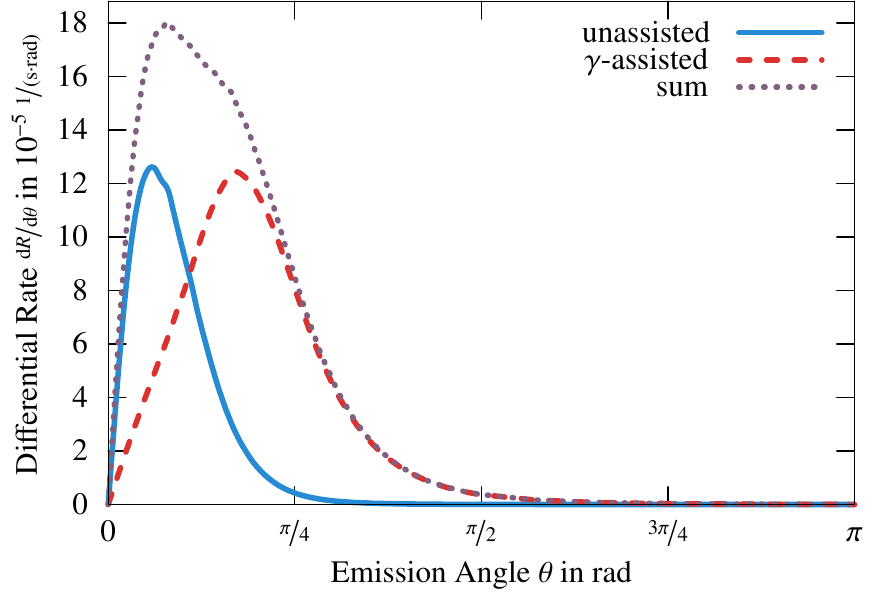}}
\\
\subfloat[][Laboratory Frame]{\figlab{spectra-lab-theta}%
\includegraphics[width=\columnwidth]{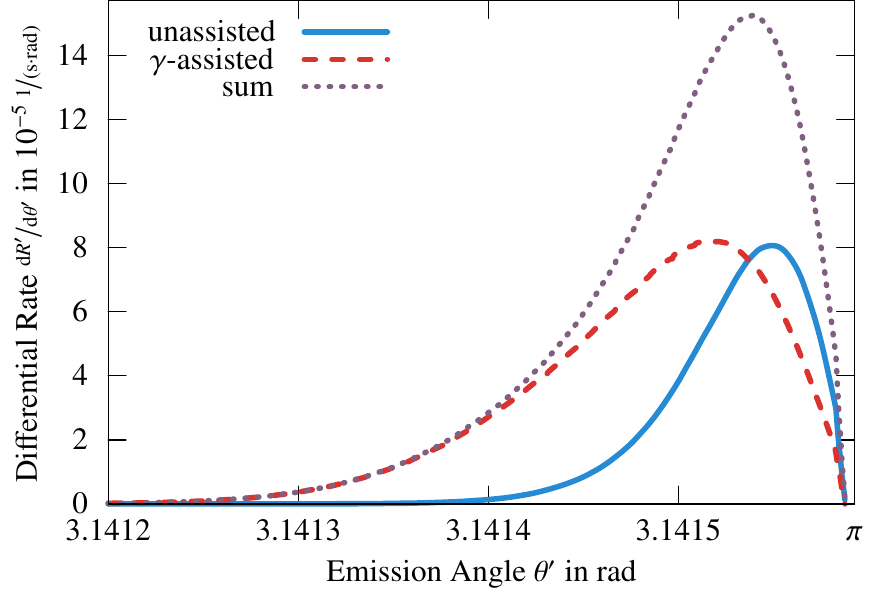}}
\caption{\figlab{spectra-theta}%
Angular-differential spectra of the unassisted and the $\gamma$-assisted process
--
The same parameters as in \subfigref{assist980} have been used at $\xi_\strong = 1.0$.
In both frames of reference, the width of the distribution is increased for the assisted case. In the nuclear rest frame, the peak of the assisted process is shifted towards larger angles. Just as in the laboratory frame, towards slightly smaller angles.
}
\end{figure}

Besides the total pair-creation rate, further insight into the differences between the assisted and the unassisted process may be found in the arising differential spectra. In particular, a potential experimental observation would rely on knowing the required acceptance of a detector and the optimal angle under which to measure.

Returning to the above parameter set with $\gamma = 7000$, where the parameter $\zeta$ (cf. \equref{param-zeta}) is relatively large, \figref{spectra-theta} shows the pair-creation rate differential in the emission angle $\theta$ in the nuclear rest frame and the laboratory frame for the both processes at $\xi_\strong = 1$, the approximate crossing point found in \subfigref{assist980}. The two key differences are the shifted peak position and the increased width of the distribution for the assisted process.

It is worth pointing out that, the minimal number of low-frequency photons to overcome the pair-creation threshold for the unassisted process in \figref{spectra-theta} amounts to $n_2^\text{min} = 45$, whereas the assisted process requires only $n_2^\text{min} = 15$. The difference between these threshold numbers corresponds to the frequency ratio $\nicefrac{\omega_1}{\omega_2}$. The main contribution to the total pair-creation rates, though, stems from substantially larger photon numbers. For the unassisted process, the typical number $n_2$ ranges from about $50$ to $75$, whereas it ranges from about $18$ to $50$ for the assisted process. Hence, the broader angular distribution arises in the case where the total number of absorbed photons is smaller.
Interestingly, a similar effect of angular broadening for smaller photon numbers has also been found for pair creation by few-photon absorption in the perturbative interaction domain \cite{smueller}. There, however, a small shift of the emission maximum towards larger angles has been found for larger photon numbers, in contrast to the present situation.

Note that, the Lorentz transformation contracts the full-$\pi$ spectrum of the nuclear rest frame (\subfigref{spectra-nuc-theta}) into a small range in the laboratory frame (\subfigref{spectra-lab-theta}). Furthermore, it maps small angles to large angles and vice versa. Therefore, when comparing the distributions in one reference frame with its counterpart in the other, the spectra show a mirrored behavior.
To both spectra, the summed-up differential rates $\nicefrac{\diff R}{\diff \theta^{(\prime)}}$, with $R = R_0 + R_1$ as defined in \equref{rate-sum-n1}, have been added. Note that, the rate of the unassisted process could in principle be measured independently by turning off the assisting light source. In contrast, the assisted process occurs only simultaneously with its unassisted counterpart. Nevertheless, the distinction between the two processes in their contribution to the summed-up rate is more illustrative.

\begin{figure}[t]
\centering
\subfloat[][Nuclear Rest Frame]{\figlab{spectra-nuc-energy}%
\includegraphics[width=\columnwidth]{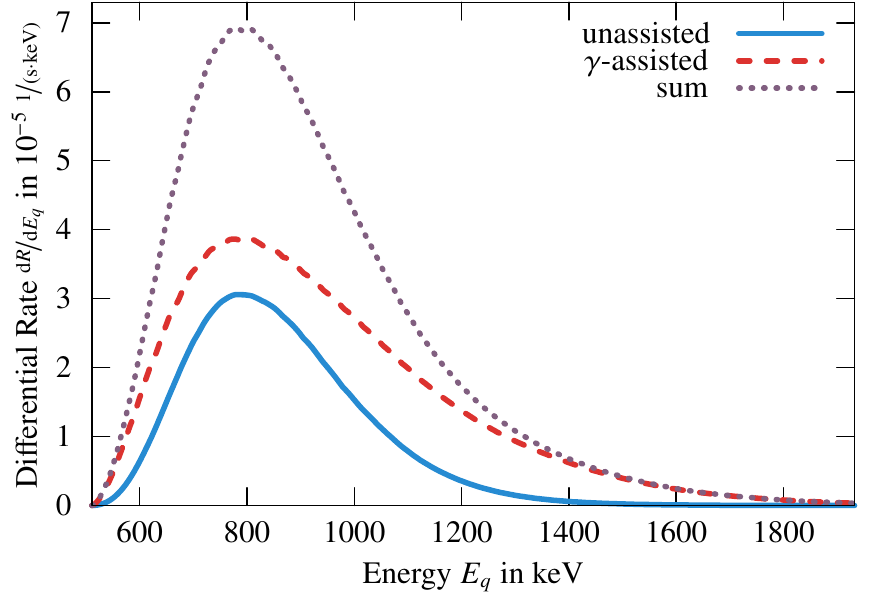}}
\\
\subfloat[][Laboratory Frame]{\figlab{spectra-lab-energy}%
\includegraphics[width=\columnwidth]{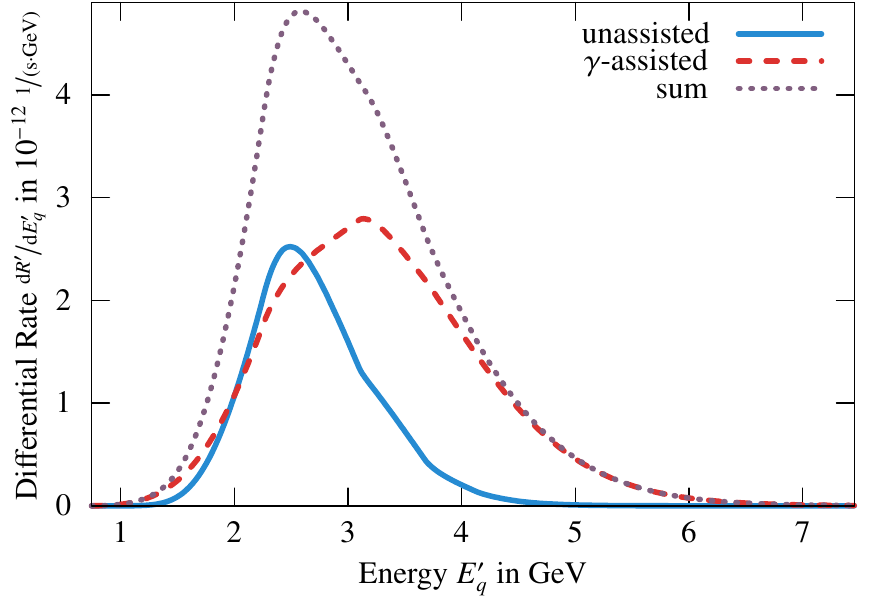}}
\caption{\figlab{spectra-energy}%
Energy-differential spectra of the unassisted and the $\gamma$-assisted process
--
The same parameters as in \subfigref{assist980} have been used at $\xi_\strong = 1.0$.
In both frames of reference, the width of the distribution is increased for the assisted case. In the nuclear rest frame the peak position remains unchanged, while it shifts towards larger energies in the laboratory frame.
}
\end{figure}

The energy distribution is depicted in \figref{spectra-energy} for the same parameters as above. Similar to \figref{spectra-theta}, a broadening for the assisted process is found in both frames of reference. However, only in the laboratory frame the peak position is shifted.
The latter may be attributed to the interweaving of the emission angle as given in \subfigref{spectra-nuc-theta} (contained in the $z$-component of the four-momentum) and the energy as given in \subfigref{spectra-nuc-energy} 
by the Lorentz transformation to the laboratory frame:
\begin{equation}
E'_q = \gamma \left( E_q + \beta \, q \cos(\theta) \right).
\end{equation}
Hence, the peak position shift in the former manifests itself in the Lorentz-transformed version of the latter.
Again, the summed-up rates $\nicefrac{\diff R}{\diff E^{(\prime)}_q}$ have been added to both spectra for the same reason as stated above.

\subsection{Doubly \texorpdfstring{$\gamma$}{Gamma}-Assisted Process}

\begin{figure}[t]
\centering
\includegraphics[width=\columnwidth]{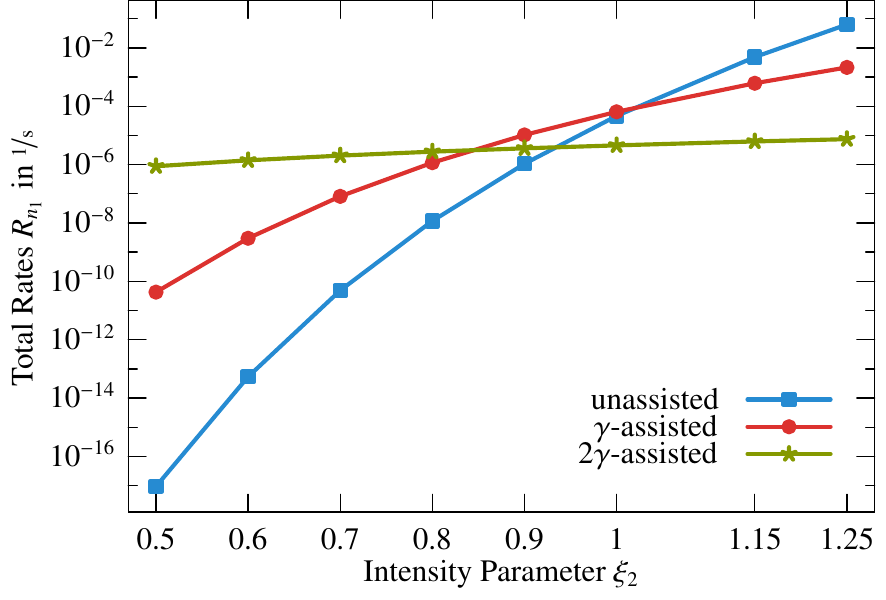}
\caption{\figlab{assist-n2}%
Total pair-creation rates of the unassisted, the singly $\gamma$-assisted, and the doubly $\gamma$-assisted process in the nuclear rest frame
--
With the parameters 
$\omega_\weak = 490 \unit{keV}$, $\omega_\strong = 32.62 \unit{keV}$, and $\xi_\weak = \sci{1.45}{-8}$.
For the smallest $\xi_\strong$, the doubly assisted process dominates. In an intermediate range, $0.85 \lesssim \xi_\strong \lesssim 1.02$, the singly assisted process is strongest. Just as the unassisted process, for the highest $\xi_\strong$.
}
\end{figure}

Extending our results to a higher-order assisted process may be straightforwardly done by halving $\omega_\weak$, the frequency of the assisting mode. This way we may compare the unassisted case (which obviously remains unchanged) to the assisted processes with one or two photons of high energy.
Furthermore, the intensity of the assisting mode is increased to $\xi_\weak = \sci{5}{-5}$ to ensure better comparability of the three processes of interest.
Besides, the parameters are identical to the above case of $\gamma = 7000$.

Analogous to \figref{assist}, the total rates for these three processes are shown in \figref{assist-n2} as function of $\xi_\strong$. We note that, in agreement with the above found behavior, for the lowest values of $\xi_\strong$ the highest number of assisting photons, two, is dominant. From $\xi_\strong \approx 0.85$ onwards, the singly assisted process overgrows the former, which is in turn overgrown by the unassisted case for the highest depicted intensities $(\xi_\strong \gtrsim 1.02)$.%
\footnote{%
Here, pair-creation by the weak mode ($\xi_\weak = \sci{5}{-5}$, $\omega_\weak = 490 \unit{keV}$) alone is dominated by the absorption of three $\omega_\weak$-photons, with a rate of $\sci{2.6}{-12} \nicefrac{1}{\text{s}}$.
%
} %
It is worth pointing out that here, the singly $\gamma$-assisted process with $\omega_\weak = 490 \unit{keV}$ represents the aforementioned case where the parameter $\delta$ (cf. \equref{param-delta}) is larger than unity, as $\delta \approx 3.4$.

\begin{figure}[t]
\centering
\subfloat[][]{\figlab{spectra-theta-n2}%
\includegraphics[width=\columnwidth]{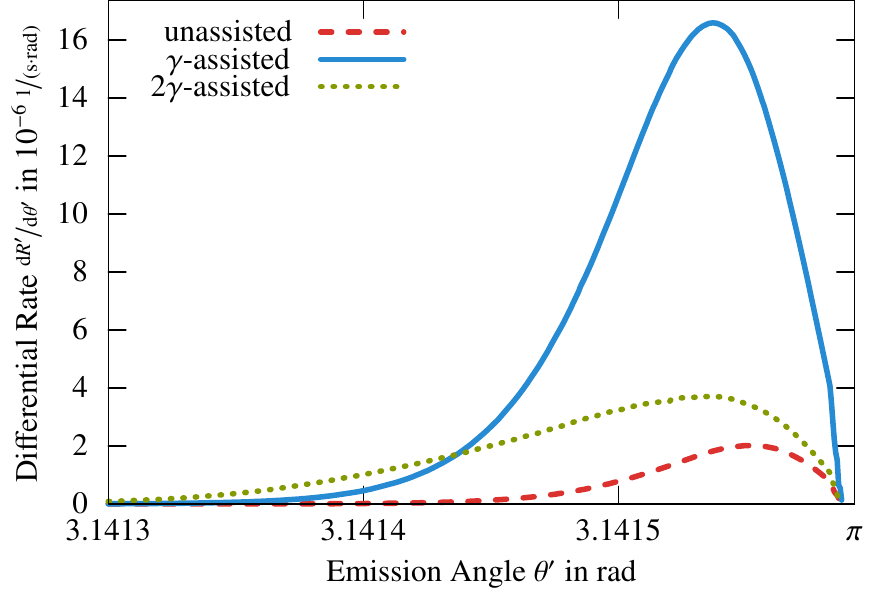}}
\\
\subfloat[][]{\figlab{spectra-energy-n2}%
\includegraphics[width=\columnwidth]{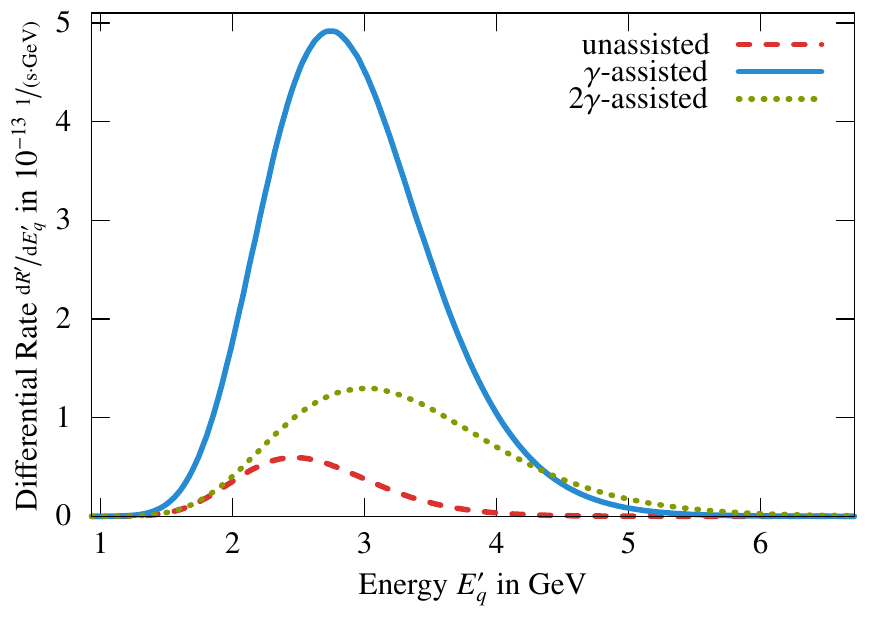}}
\caption{\figlab{spectra-n2}%
Differential spectra of the unassisted, the singly $\gamma$-assisted, and the doubly $\gamma$-assisted process in the laboratory frame
--
The same parameters as in \figref{assist-n2} have been used at $\xi_\strong = 0.9$.
In accordance to Figs.~\ref{fig:spectra-theta} and \ref{fig:spectra-energy}, the width of both spectra is larger the more assisting photons are involved. Similarly, shifts of the peak position occur.
}
\end{figure}

A similar agreement with the behavior found in Figs.~\ref{fig:spectra-theta} and \ref{fig:spectra-energy} is visible for the angular and energy spectra depicted in \figref{spectra-n2} for $\xi_\strong = 0.9$, where the three processes are of approximately equal strength. Evidently, the distributions are further widened and the peak position is shifted for the process with two assisting photons.

\section{Summary and Conclusion} \seclab{concl}

We have studied the strong-field Bethe--Heitler process of electron-positron pair creation in relativistic proton-laser collisions, with the laser field consisting of a strong low-frequency and a weak high-frequency mode. Focusing on the nonperturbative interaction regime, where the intensity parameter of the strong mode is in the order of unity $(\xi_\strong \approx 1)$, we have shown that the assistance of the high-frequency field component can largely enhance the total pair-creation rate. The latter still exhibits a manifestly nonperturbative field dependence, similar to the famous Schwinger rate. This way we extended previous results valid for $\xi_\strong \gg 1$ \cite{dipiazza-tunnel,dipiazza} to the intermediate laser-matter coupling regime and showed, moreover, that the field scaling of the Schwinger-like exponent changes significantly when the high laser frequency closely approaches the pair-creation threshold (in the proton rest frame).

Additionally, we have demonstrated that the $\gamma$-assisted pair-creation mechanism exerts a characteristic influence on the angular and energy distributions of created particles. It leads to shifts of the emission maxima and a broadening of the distributions.
Finally, a regime of parameters has been identified where the assistance of two high-frequency photons strongly affects, or even dominates, the pair-creation rate. In accordance with the above, characteristic changes were also found in the underlying spectra.

Our results can be tested experimentally by combining the LHC proton beam with a counterpropagating laser pulse comprised of a highly intense optical mode and a high-frequency (XUV or soft X-ray) mode of moderate intensity.

\section*{Acknowledgments}

Funding for this project by the German Research Foundation (DFG) under Grant No. MU 3149/1-1 is gratefully acknowledged.
S.\,A. also wishes to thank the Heidelberg Graduate School of Fundamental Physics (HGSFP) for the generous travel support.

\section*{References}
\bibliography{b2}

\begin{thebibliography}{47}
\expandafter\ifx\csname natexlab\endcsname\relax\def\natexlab#1{#1}\fi
\providecommand{\url}[1]{\texttt{#1}}
\providecommand{\href}[2]{#2}
\providecommand{\path}[1]{#1}
\providecommand{\DOIprefix}{doi:}
\providecommand{\ArXivprefix}{arXiv:}
\providecommand{\URLprefix}{URL: }
\providecommand{\Pubmedprefix}{pmid:}
\providecommand{\doi}[1]{\href{http://dx.doi.org/#1}{\path{#1}}}
\providecommand{\Pubmed}[1]{\href{pmid:#1}{\path{#1}}}
\providecommand{\bibinfo}[2]{#2}
\ifx\xfnm\relax \def\xfnm[#1]{\unskip,\space#1}\fi
\bibitem[{Sauter(1931)}]{Sauter}
\bibinfo{author}{F.~Sauter},
\newblock \bibinfo{title}{\"{U}ber das {V}erhalten eines {E}lektrons im
  homogenen elektrischen {F}eld nach der relativistischen {T}heorie {D}iracs},
\newblock \bibinfo{journal}{Z. Physik} \bibinfo{volume}{69}
  (\bibinfo{year}{1931}) \bibinfo{pages}{742--764}.
\bibitem[{Heisenberg and Euler(1936)}]{heisenberg}
\bibinfo{author}{W.~Heisenberg}, \bibinfo{author}{H.~Euler},
\newblock \bibinfo{title}{Folgerungen aus der {D}iracschen {T}heorie des
  {P}ositrons},
\newblock \bibinfo{journal}{Z. Physik} \bibinfo{volume}{98}
  (\bibinfo{year}{1936}) \bibinfo{pages}{714--732}.
\bibitem[{Greiner et~al.(1985)Greiner, M{\"u}ller, and Rafelski}]{Greiner}
\bibinfo{author}{W.~Greiner}, \bibinfo{author}{B.~M{\"u}ller},
  \bibinfo{author}{J.~Rafelski}, \bibinfo{title}{Quantum {E}lectrodynamics of
  {S}trong {F}ields}, \bibinfo{publisher}{Springer}, \bibinfo{address}{Berlin,
  Heidelberg}, \bibinfo{year}{1985}.
\bibitem[{Schwinger(1951)}]{Schwinger}
\bibinfo{author}{J.~Schwinger},
\newblock \bibinfo{title}{On gauge invariance and vacuum polarization},
\newblock \bibinfo{journal}{Phys. Rev.} \bibinfo{volume}{82}
  (\bibinfo{year}{1951}) \bibinfo{pages}{664--679}.
\bibitem[{Ringwald(2001)}]{ringwald}
\bibinfo{author}{A.~Ringwald},
\newblock \bibinfo{title}{Pair production from vacuum at the focus of an
  {X}-ray free electron laser},
\newblock \bibinfo{journal}{Phys. Lett. B} \bibinfo{volume}{510}
  (\bibinfo{year}{2001}) \bibinfo{pages}{107--116}.
\bibitem[{Ruffini et~al.(2003)Ruffini, Vitagliano, and Xue}]{ruffini}
\bibinfo{author}{R.~Ruffini}, \bibinfo{author}{L.~Vitagliano},
  \bibinfo{author}{S.-S. Xue},
\newblock \bibinfo{title}{On plasma oscillations in strong electric fields},
\newblock \bibinfo{journal}{Phys. Lett. B} \bibinfo{volume}{559}
  (\bibinfo{year}{2003}) \bibinfo{pages}{12--19}.
\bibitem[{Bulanov et~al.(2010)Bulanov, Mur, Narozhny, Nees, and
  Popov}]{bulanov-schwinger}
\bibinfo{author}{S.~S. Bulanov}, \bibinfo{author}{V.~D. Mur},
  \bibinfo{author}{N.~B. Narozhny}, \bibinfo{author}{J.~Nees},
  \bibinfo{author}{V.~S. Popov},
\newblock \bibinfo{title}{Multiple colliding electromagnetic pulses: a way to
  lower the threshold of e$^+$ e$^-$ pair production from vacuum},
\newblock \bibinfo{journal}{Phys. Rev. Lett.} \bibinfo{volume}{104}
  (\bibinfo{year}{2010}) \bibinfo{pages}{220404}.
\bibitem[{Dumlu and Dunne(2010)}]{dumlu-dunne-schwinger}
\bibinfo{author}{C.~K. Dumlu}, \bibinfo{author}{G.~V. Dunne},
\newblock \bibinfo{title}{Stokes phenomenon and {S}chwinger vacuum pair
  production in time-dependent laser pulses},
\newblock \bibinfo{journal}{Phys. Rev. Lett.} \bibinfo{volume}{104}
  (\bibinfo{year}{2010}) \bibinfo{pages}{250402}.
\bibitem[{Hebenstreit et~al.(2011)Hebenstreit, Alkofer, and Gies}]{hebenstreit}
\bibinfo{author}{F.~Hebenstreit}, \bibinfo{author}{R.~Alkofer},
  \bibinfo{author}{H.~Gies},
\newblock \bibinfo{title}{Particle self-bunching in the {S}chwinger effect in
  spacetime-dependent electric fields},
\newblock \bibinfo{journal}{Phys. Rev. Lett.} \bibinfo{volume}{107}
  (\bibinfo{year}{2011}) \bibinfo{pages}{180403}.
\bibitem[{Kohlf\"urst et~al.(2013)Kohlf\"urst, Mitter, von Winckel,
  Hebenstreit, and Alkofer}]{kohlfuerst}
\bibinfo{author}{C.~Kohlf\"urst}, \bibinfo{author}{M.~Mitter},
  \bibinfo{author}{G.~von Winckel}, \bibinfo{author}{F.~Hebenstreit},
  \bibinfo{author}{R.~Alkofer},
\newblock \bibinfo{title}{Optimizing the pulse shape for {S}chwinger pair
  production},
\newblock \bibinfo{journal}{Phys. Rev. D} \bibinfo{volume}{88}
  (\bibinfo{year}{2013}) \bibinfo{pages}{045028}.
\bibitem[{Ehlotzky et~al.(2009)Ehlotzky, Krajewska, and
  Kami\'nski}]{ehlotzky-review}
\bibinfo{author}{F.~Ehlotzky}, \bibinfo{author}{K.~Krajewska},
  \bibinfo{author}{J.~Z. Kami\'nski},
\newblock \bibinfo{title}{Fundamental processes of quantum electrodynamics in
  laser fields of relativistic power},
\newblock \bibinfo{journal}{Rep. Prog. Phys.} \bibinfo{volume}{72}
  (\bibinfo{year}{2009}) \bibinfo{pages}{046401}.
\bibitem[{Di~Piazza et~al.(2012)Di~Piazza, M\"uller, Hatsagortsyan, and
  Keitel}]{dipiazza-review}
\bibinfo{author}{A.~Di~Piazza}, \bibinfo{author}{C.~M\"uller},
  \bibinfo{author}{K.~Z. Hatsagortsyan}, \bibinfo{author}{C.~H. Keitel},
\newblock \bibinfo{title}{Extremely high-intensity laser interactions with
  fundamental quantum systems},
\newblock \bibinfo{journal}{Rev. Mod. Phys.} \bibinfo{volume}{84}
  (\bibinfo{year}{2012}) \bibinfo{pages}{1177--1228}.
\bibitem[{Sch\"utzhold et~al.(2008)Sch\"utzhold, Gies, and Dunne}]{schuetzhold}
\bibinfo{author}{R.~Sch\"utzhold}, \bibinfo{author}{H.~Gies},
  \bibinfo{author}{G.~Dunne},
\newblock \bibinfo{title}{Dynamically assisted {S}chwinger mechanism},
\newblock \bibinfo{journal}{Phys. Rev. Lett.} \bibinfo{volume}{101}
  (\bibinfo{year}{2008}) \bibinfo{pages}{130404}.
\bibitem[{Dunne et~al.(2009)Dunne, Gies, and Sch\"utzhold}]{dunne}
\bibinfo{author}{G.~V. Dunne}, \bibinfo{author}{H.~Gies},
  \bibinfo{author}{R.~Sch\"utzhold},
\newblock \bibinfo{title}{Catalysis of {S}chwinger vacuum pair production},
\newblock \bibinfo{journal}{Phys. Rev. D} \bibinfo{volume}{80}
  (\bibinfo{year}{2009}) \bibinfo{pages}{111301}.
\bibitem[{Orthaber et~al.(2011)Orthaber, Hebenstreit, and Alkofer}]{orthaber}
\bibinfo{author}{M.~Orthaber}, \bibinfo{author}{F.~Hebenstreit},
  \bibinfo{author}{R.~Alkofer},
\newblock \bibinfo{title}{Momentum spectra for dynamically assisted {S}chwinger
  pair production},
\newblock \bibinfo{journal}{Phys. Lett. B} \bibinfo{volume}{698}
  (\bibinfo{year}{2011}) \bibinfo{pages}{80--85}.
\bibitem[{Fey and Sch\"utzhold(2012)}]{fey}
\bibinfo{author}{C.~Fey}, \bibinfo{author}{R.~Sch\"utzhold},
\newblock \bibinfo{title}{Momentum dependence in the dynamically assisted
  {S}auter--{S}chwinger effect},
\newblock \bibinfo{journal}{Phys. Rev. D} \bibinfo{volume}{85}
  (\bibinfo{year}{2012}) \bibinfo{pages}{025004}.
\bibitem[{Li et~al.(2014)Li, Lu, and Xie}]{li-zi-liang-boson}
\bibinfo{author}{Z.-L. Li}, \bibinfo{author}{D.~Lu}, \bibinfo{author}{B.-S.
  Xie},
\newblock \bibinfo{title}{Multiple-slit interference effect in the time domain
  for boson pair production},
\newblock \bibinfo{journal}{Phys. Rev. D} \bibinfo{volume}{89}
  (\bibinfo{year}{2014}) \bibinfo{pages}{067701}.
\bibitem[{M\"uller et~al.(2003)M\"uller, Voitkiv, and Gr\"un}]{cm-circ}
\bibinfo{author}{C.~M\"uller}, \bibinfo{author}{A.~B. Voitkiv},
  \bibinfo{author}{N.~Gr\"un},
\newblock \bibinfo{title}{Differential rates for multiphoton pair production by
  an ultrarelativistic nucleus colliding with an intense laser beam},
\newblock \bibinfo{journal}{Phys. Rev. A} \bibinfo{volume}{67}
  (\bibinfo{year}{2003}) \bibinfo{pages}{063407}.
\bibitem[{Avetissian et~al.(2003)Avetissian, Avetissian, Mkrtchian, and
  Sedrakian}]{Avetissian}
\bibinfo{author}{H.~K. Avetissian}, \bibinfo{author}{A.~K. Avetissian},
  \bibinfo{author}{G.~F. Mkrtchian}, \bibinfo{author}{K.~V. Sedrakian},
\newblock \bibinfo{title}{X-ray free electron laser for electron-positron pair
  production on the nuclei},
\newblock \bibinfo{journal}{Nucl. Instrum. Methods Phys. Res. Sect. A}
  \bibinfo{volume}{507} (\bibinfo{year}{2003}) \bibinfo{pages}{582 -- 586}.
\bibitem[{Sieczka et~al.(2006)Sieczka, Krajewska, Kami\'nski, Panek, and
  Ehlotzky}]{Sieczka}
\bibinfo{author}{P.~Sieczka}, \bibinfo{author}{K.~Krajewska},
  \bibinfo{author}{J.~Z. Kami\'nski}, \bibinfo{author}{P.~Panek},
  \bibinfo{author}{F.~Ehlotzky},
\newblock \bibinfo{title}{Electron-positron pair creation by powerful laser-ion
  impact},
\newblock \bibinfo{journal}{Phys. Rev. A} \bibinfo{volume}{73}
  (\bibinfo{year}{2006}) \bibinfo{pages}{053409}.
\bibitem[{Milstein et~al.(2006)Milstein, M\"uller, Hatsagortsyan, Jentschura,
  and Keitel}]{Milstein}
\bibinfo{author}{A.~I. Milstein}, \bibinfo{author}{C.~M\"uller},
  \bibinfo{author}{K.~Z. Hatsagortsyan}, \bibinfo{author}{U.~D. Jentschura},
  \bibinfo{author}{C.~H. Keitel},
\newblock \bibinfo{title}{Polarization-operator approach to electron-positron
  pair production in combined laser and {C}oulomb fields},
\newblock \bibinfo{journal}{Phys. Rev. A} \bibinfo{volume}{73}
  (\bibinfo{year}{2006}) \bibinfo{pages}{062106}.
\bibitem[{Kuchiev and Robinson(2007)}]{Kuchiev}
\bibinfo{author}{M.~Y. Kuchiev}, \bibinfo{author}{D.~J. Robinson},
\newblock \bibinfo{title}{Electron-positron pair creation by {C}oulomb and
  laser fields in the tunneling regime},
\newblock \bibinfo{journal}{Phys. Rev. A} \bibinfo{volume}{76}
  (\bibinfo{year}{2007}) \bibinfo{pages}{012107}.
\bibitem[{M\"uller and M\"uller(2011)}]{tomueller2}
\bibinfo{author}{T.-O. M\"uller}, \bibinfo{author}{C.~M\"uller},
\newblock \bibinfo{title}{Spin correlations in nonperturbative
  electron--positron pair creation by petawatt laser pulses colliding with a
  {T}e{V} proton beam},
\newblock \bibinfo{journal}{Physics Letters B} \bibinfo{volume}{696}
  (\bibinfo{year}{2011}) \bibinfo{pages}{201--206}.
\bibitem[{Fillion-Gourdeau et~al.(2013)Fillion-Gourdeau, Lorin, and
  Bandrauk}]{fillion}
\bibinfo{author}{F.~Fillion-Gourdeau}, \bibinfo{author}{E.~Lorin},
  \bibinfo{author}{A.~D. Bandrauk},
\newblock \bibinfo{title}{Resonantly enhanced pair production in a simple
  diatomic model},
\newblock \bibinfo{journal}{Phys. Rev. Lett.} \bibinfo{volume}{110}
  (\bibinfo{year}{2013}) \bibinfo{pages}{013002}.
\bibitem[{Burke et~al.(1997)Burke, Field, Horton-Smith, Spencer, Walz,
  Berridge, Bugg, Shmakov, Weidemann, Bula, McDonald, Prebys, Bamber, Boege,
  Koffas, Kotseroglou, Melissinos, Meyerhofer, Reis, and Ragg}]{e144-1997}
\bibinfo{author}{D.~L. Burke}, \bibinfo{author}{R.~C. Field},
  \bibinfo{author}{G.~Horton-Smith}, \bibinfo{author}{J.~E. Spencer},
  \bibinfo{author}{D.~Walz}, \bibinfo{author}{S.~C. Berridge},
  \bibinfo{author}{W.~M. Bugg}, \bibinfo{author}{K.~Shmakov},
  \bibinfo{author}{A.~W. Weidemann}, \bibinfo{author}{C.~Bula},
  \bibinfo{author}{K.~T. McDonald}, \bibinfo{author}{E.~J. Prebys},
  \bibinfo{author}{C.~Bamber}, \bibinfo{author}{S.~J. Boege},
  \bibinfo{author}{T.~Koffas}, \bibinfo{author}{T.~Kotseroglou},
  \bibinfo{author}{A.~C. Melissinos}, \bibinfo{author}{D.~D. Meyerhofer},
  \bibinfo{author}{D.~A. Reis}, \bibinfo{author}{W.~Ragg},
\newblock \bibinfo{title}{Positron production in multiphoton light-by-light
  scattering},
\newblock \bibinfo{journal}{Phys. Rev. Lett.} \bibinfo{volume}{79}
  (\bibinfo{year}{1997}) \bibinfo{pages}{1626--1629}.
\bibitem[{Reiss(2009)}]{reiss-trident}
\bibinfo{author}{H.~R. Reiss},
\newblock \bibinfo{title}{Special analytical properties of ultrastrong coherent
  fields},
\newblock \bibinfo{journal}{The European Physical Journal D}
  \bibinfo{volume}{55} (\bibinfo{year}{2009}) \bibinfo{pages}{365--374}.
\bibitem[{Hu et~al.(2010)Hu, M\"uller, and Keitel}]{hu-trident}
\bibinfo{author}{H.~Hu}, \bibinfo{author}{C.~M\"uller}, \bibinfo{author}{C.~H.
  Keitel},
\newblock \bibinfo{title}{Complete {QED} theory of multiphoton trident pair
  production in strong laser fields},
\newblock \bibinfo{journal}{Phys. Rev. Lett.} \bibinfo{volume}{105}
  (\bibinfo{year}{2010}) \bibinfo{pages}{080401}.
\bibitem[{Di~Piazza et~al.(2009)Di~Piazza, L\"otstedt, Milstein, and
  Keitel}]{dipiazza-tunnel}
\bibinfo{author}{A.~Di~Piazza}, \bibinfo{author}{E.~L\"otstedt},
  \bibinfo{author}{A.~I. Milstein}, \bibinfo{author}{C.~H. Keitel},
\newblock \bibinfo{title}{Barrier control in tunneling e$^+$-e$^-$
  photoproduction},
\newblock \bibinfo{journal}{Phys. Rev. Lett.} \bibinfo{volume}{103}
  (\bibinfo{year}{2009}) \bibinfo{pages}{170403}.
\bibitem[{Di~Piazza et~al.(2010)Di~Piazza, L\"otstedt, Milstein, and
  Keitel}]{dipiazza}
\bibinfo{author}{A.~Di~Piazza}, \bibinfo{author}{E.~L\"otstedt},
  \bibinfo{author}{A.~I. Milstein}, \bibinfo{author}{C.~H. Keitel},
\newblock \bibinfo{title}{Effect of a strong laser field on electron-positron
  photoproduction by relativistic nuclei},
\newblock \bibinfo{journal}{Phys. Rev. A} \bibinfo{volume}{81}
  (\bibinfo{year}{2010}) \bibinfo{pages}{062122}.
\bibitem[{Jansen and M\"uller(2013)}]{Jansen}
\bibinfo{author}{M.~J.~A. Jansen}, \bibinfo{author}{C.~M\"uller},
\newblock \bibinfo{title}{Strongly enhanced pair production in combined high-
  and low-frequency laser fields},
\newblock \bibinfo{journal}{Phys. Rev. A} \bibinfo{volume}{88}
  (\bibinfo{year}{2013}) \bibinfo{pages}{052125}.
\bibitem[{Jiang et~al.(2012)Jiang, Su, Lv, Lu, Li, Grobe, and Su}]{jiang-grobe}
\bibinfo{author}{M.~Jiang}, \bibinfo{author}{W.~Su}, \bibinfo{author}{Z.~Q.
  Lv}, \bibinfo{author}{X.~Lu}, \bibinfo{author}{Y.~J. Li},
  \bibinfo{author}{R.~Grobe}, \bibinfo{author}{Q.~Su},
\newblock \bibinfo{title}{Pair creation enhancement due to combined external
  fields},
\newblock \bibinfo{journal}{Phys. Rev. A} \bibinfo{volume}{85}
  (\bibinfo{year}{2012}) \bibinfo{pages}{033408}.
\bibitem[{Jiang et~al.(2013)Jiang, Lv, Sheng, Grobe, and Su}]{jiang-grobe2}
\bibinfo{author}{M.~Jiang}, \bibinfo{author}{Q.~Z. Lv}, \bibinfo{author}{Z.~M.
  Sheng}, \bibinfo{author}{R.~Grobe}, \bibinfo{author}{Q.~Su},
\newblock \bibinfo{title}{Enhancement of electron-positron pair creation due to
  transient excitation of field-induced bound states},
\newblock \bibinfo{journal}{Phys. Rev. A} \bibinfo{volume}{87}
  (\bibinfo{year}{2013}) \bibinfo{pages}{042503}.
\bibitem[{Krajewska and Kami\'nski(2012)}]{krajewska-phase}
\bibinfo{author}{K.~Krajewska}, \bibinfo{author}{J.~Z. Kami\'nski},
\newblock \bibinfo{title}{Phase effects in laser-induced electron-positron pair
  creation},
\newblock \bibinfo{journal}{Phys. Rev. A} \bibinfo{volume}{85}
  (\bibinfo{year}{2012}) \bibinfo{pages}{043404}.
\bibitem[{Roshchupkin(2001)}]{Roshchupkin}
\bibinfo{author}{S.~P. Roshchupkin},
\newblock \bibinfo{title}{Interference effect in the photoproduction of
  electron-positron pairs on a nucleus in the field of two light waves},
\newblock \bibinfo{journal}{Phys. At. Nucl.} \bibinfo{volume}{64}
  (\bibinfo{year}{2001}) \bibinfo{pages}{243--252}.
\bibitem[{Augustin and M\"uller(2013)}]{augustin}
\bibinfo{author}{S.~Augustin}, \bibinfo{author}{C.~M\"uller},
\newblock \bibinfo{title}{Interference effects in {B}ethe--{H}eitler pair
  creation in a bichromatic laser field},
\newblock \bibinfo{journal}{Phys. Rev. A} \bibinfo{volume}{88}
  (\bibinfo{year}{2013}) \bibinfo{pages}{022109}.
\bibitem[{Augustin and M\"uller(2014)}]{augustin2}
\bibinfo{author}{S.~Augustin}, \bibinfo{author}{C.~M\"uller},
\newblock \bibinfo{title}{Nonlinear {B}ethe--{H}eitler pair creation in an
  intense two-mode laser field},
\newblock \bibinfo{journal}{J. Phys. Conf. Ser.} \bibinfo{volume}{497}
  (\bibinfo{year}{2014}) \bibinfo{pages}{012020}.
\bibitem[{Allor et~al.(2008)Allor, Cohen, and McGady}]{allor-graphene}
\bibinfo{author}{D.~Allor}, \bibinfo{author}{T.~D. Cohen},
  \bibinfo{author}{D.~A. McGady},
\newblock \bibinfo{title}{Schwinger mechanism and graphene},
\newblock \bibinfo{journal}{Phys. Rev. D} \bibinfo{volume}{78}
  (\bibinfo{year}{2008}) \bibinfo{pages}{096009}.
\bibitem[{Klimchitskaya and Mostepanenko(2013)}]{klimchitskaya-graphene}
\bibinfo{author}{G.~L. Klimchitskaya}, \bibinfo{author}{V.~M. Mostepanenko},
\newblock \bibinfo{title}{Creation of quasiparticles in graphene by a
  time-dependent electric field},
\newblock \bibinfo{journal}{Phys. Rev. D} \bibinfo{volume}{87}
  (\bibinfo{year}{2013}) \bibinfo{pages}{125011}.
\bibitem[{Szpak and Sch\"utzhold(2012)}]{szpak-cold2}
\bibinfo{author}{N.~Szpak}, \bibinfo{author}{R.~Sch\"utzhold},
\newblock \bibinfo{title}{Optical lattice quantum simulator for quantum
  electrodynamics in strong external fields: spontaneous pair creation and the
  {S}auter--{S}chwinger effect},
\newblock \bibinfo{journal}{New Journal of Physics} \bibinfo{volume}{14}
  (\bibinfo{year}{2012}) \bibinfo{pages}{035001}.
\bibitem[{Dreisow et~al.(2012)Dreisow, Longhi, Nolte, T\"unnermann, and
  Szameit}]{dreisow-waveguide}
\bibinfo{author}{F.~Dreisow}, \bibinfo{author}{S.~Longhi},
  \bibinfo{author}{S.~Nolte}, \bibinfo{author}{A.~T\"unnermann},
  \bibinfo{author}{A.~Szameit},
\newblock \bibinfo{title}{Vacuum instability and pair production in an optical
  setting},
\newblock \bibinfo{journal}{Phys. Rev. Lett.} \bibinfo{volume}{109}
  (\bibinfo{year}{2012}) \bibinfo{pages}{110401}.
\bibitem[{Bjorken and Drell(1964)}]{bjorken-drell}
\bibinfo{author}{J.~D. Bjorken}, \bibinfo{author}{S.~D. Drell},
  \bibinfo{title}{Relativistic {Q}uantum {M}echanics},
  \bibinfo{publisher}{McGraw-Hill}, \bibinfo{address}{New York},
  \bibinfo{year}{1964}. \bibinfo{note}{In Sec. 7.1}.
\bibitem[{Zhang et~al.(2007)Zhang, Lytle, Popmintchev, Zhou, Kapteyn, Murnane,
  and Cohen}]{zhang-hhg}
\bibinfo{author}{X.~Zhang}, \bibinfo{author}{A.~L. Lytle},
  \bibinfo{author}{T.~Popmintchev}, \bibinfo{author}{X.~Zhou},
  \bibinfo{author}{H.~C. Kapteyn}, \bibinfo{author}{M.~M. Murnane},
  \bibinfo{author}{O.~Cohen},
\newblock \bibinfo{title}{Quasi-phase-matching and quantum-path control of
  high-harmonic generation using counterpropagating light},
\newblock \bibinfo{journal}{Nat Phys} \bibinfo{volume}{3}
  (\bibinfo{year}{2007}) \bibinfo{pages}{270--275}.
\bibitem[{Ackermann et~al.(2007)Ackermann, Asova, Ayvazyan, Azima, Baboi, Bahr,
  Balandin, Beutner, Brandt, Bolzmann, Brinkmann, Brovko, Castellano, Castro,
  Catani, Chiadroni, Choroba, Cianchi, Costello, and Cubaynes}]{ackermann-fel}
\bibinfo{author}{W.~Ackermann}, \bibinfo{author}{G.~Asova},
  \bibinfo{author}{V.~Ayvazyan}, \bibinfo{author}{A.~Azima},
  \bibinfo{author}{N.~Baboi}, \bibinfo{author}{J.~Bahr},
  \bibinfo{author}{V.~Balandin}, \bibinfo{author}{B.~Beutner},
  \bibinfo{author}{A.~Brandt}, \bibinfo{author}{A.~Bolzmann},
  \bibinfo{author}{R.~Brinkmann}, \bibinfo{author}{O.~I. Brovko},
  \bibinfo{author}{M.~Castellano}, \bibinfo{author}{P.~Castro},
  \bibinfo{author}{L.~Catani}, \bibinfo{author}{E.~Chiadroni},
  \bibinfo{author}{S.~Choroba}, \bibinfo{author}{A.~Cianchi},
  \bibinfo{author}{J.~T. Costello}, \bibinfo{author}{D.~Cubaynes},
\newblock \bibinfo{title}{Operation of a free-electron laser from the extreme
  ultraviolet to the water window},
\newblock \bibinfo{journal}{Nat. Photon.} \bibinfo{volume}{1}
  (\bibinfo{year}{2007}) \bibinfo{pages}{336--342}.
\bibitem[{Br\"uning et~al.(2004)Br\"uning, Collier, Lebrun, Myers, Ostojic,
  Poole, and Proudlock}]{LHC-TDR}
\bibinfo{author}{O.~S. Br\"uning}, \bibinfo{author}{P.~Collier},
  \bibinfo{author}{P.~Lebrun}, \bibinfo{author}{S.~Myers},
  \bibinfo{author}{R.~Ostojic}, \bibinfo{author}{J.~Poole},
  \bibinfo{author}{P.~Proudlock},
\newblock \bibinfo{title}{The {LHC} {M}ain {R}ing},
\newblock \bibinfo{journal}{LHC Design Report} \bibinfo{volume}{I}
  (\bibinfo{year}{2004}).
\bibitem[{Mashiko et~al.(2006)Mashiko, Suda, and
  Midorikawa}]{mashiko-hhg-intense}
\bibinfo{author}{H.~Mashiko}, \bibinfo{author}{A.~Suda},
  \bibinfo{author}{K.~Midorikawa},
\newblock \bibinfo{title}{Focusing multiple high-order harmonics in the
  extreme-ultraviolet and soft-x-ray regions by a platinum-coated ellipsoidal
  mirror},
\newblock \bibinfo{journal}{Appl. Opt.} \bibinfo{volume}{45}
  (\bibinfo{year}{2006}) \bibinfo{pages}{573--577}.
\bibitem[{Young et~al.(2010)Young, Kanter, Kr\"assig, Li, March, Pratt, Santra,
  Southworth, Rohringer, DiMauro, Doumy, Roedig, Berrah, Fang, Hoener,
  Bucksbaum, Cryan, Ghimire, Glownia, Reis, Bozek, Bostedt, and
  Messerschmidt}]{young-fel-intense}
\bibinfo{author}{L.~Young}, \bibinfo{author}{E.~P. Kanter},
  \bibinfo{author}{B.~Kr\"assig}, \bibinfo{author}{Y.~Li},
  \bibinfo{author}{A.~M. March}, \bibinfo{author}{S.~T. Pratt},
  \bibinfo{author}{R.~Santra}, \bibinfo{author}{S.~H. Southworth},
  \bibinfo{author}{N.~Rohringer}, \bibinfo{author}{L.~F. DiMauro},
  \bibinfo{author}{G.~Doumy}, \bibinfo{author}{C.~A. Roedig},
  \bibinfo{author}{N.~Berrah}, \bibinfo{author}{L.~Fang},
  \bibinfo{author}{M.~Hoener}, \bibinfo{author}{P.~H. Bucksbaum},
  \bibinfo{author}{J.~P. Cryan}, \bibinfo{author}{S.~Ghimire},
  \bibinfo{author}{J.~M. Glownia}, \bibinfo{author}{D.~A. Reis},
  \bibinfo{author}{J.~D. Bozek}, \bibinfo{author}{C.~Bostedt},
  \bibinfo{author}{M.~Messerschmidt},
\newblock \bibinfo{title}{Femtosecond electronic response of atoms to
  ultra-intense {X}-rays},
\newblock \bibinfo{journal}{Nature} \bibinfo{volume}{466}
  (\bibinfo{year}{2010}) \bibinfo{pages}{56--61}.
\bibitem[{M\"uller and M\"uller(2009)}]{smueller}
\bibinfo{author}{S.~J. M\"uller}, \bibinfo{author}{C.~M\"uller},
\newblock \bibinfo{title}{Few-photon electron-positron pair creation by
  relativistic muon impact on intense laser beams},
\newblock \bibinfo{journal}{Phys. Rev. D} \bibinfo{volume}{80}
  (\bibinfo{year}{2009}) \bibinfo{pages}{053014}.

\end{thebibliography}

\end{document}